\documentclass[fleqn,usenatbib]{mnras}

\usepackage{newtxtext,newtxmath}

\usepackage[T1]{fontenc}

\DeclareRobustCommand{\VAN}[3]{#2}
\let\VANthebibliography\thebibliography
\def\thebibliography{\DeclareRobustCommand{\VAN}[3]{##3}\VANthebibliography}

\usepackage{graphicx}	
\usepackage{amsmath}	

\title[GRB 221009A in Fermi data]{GRB 221009A, its precursor and two afterglows in the Fermi data}

\author[B. Stern and I. Tkachev]{
Boris E. Stern,$^{1}$\thanks{E-mail: stern@inr.ru}
I.I. Tkachev,$^{1,2}$
\\
$^{1}$Institute for Nuclear Research of the Russian Academy of Sciences, Moscow 117312, Russia\\
$^{2}$Physics Department and Laboratory of Cosmology and Elementary Particle Physics, Novosibirsk State University, Novosibirsk 630090, Russia\\
}

\pubyear{2023}

\begin{document}
\label{firstpage}
\pagerange{\pageref{firstpage}--\pageref{lastpage}}
\maketitle

\begin{abstract}
We study  GRB 221009A,  the brightest gamma-ray burst  in the history of observations, using  {\it Fermi} data. To calibrate them for large inclination angles,  we use the Vela X gamma-ray source. Light curves in different spectral ranges demonstrate a 300\,s overlap of afterglow and delayed episodes of soft prompt emission.  We demonstrate that a relatively weak burst precursor that occurs 3 minutes before the main episode has its own afterglow, i.e., presumably, its own external shock. The main afterglow is  the brightest one, includes a photon with an energy of 400 GeV 9 hours after the burst, and is visible in the LAT data for up to two days.
\end{abstract}



\section{Introduction}

The recent and   brightest gamma-ray burst, GRB 221009A, has been detected by many space X-ray -- $\gamma$-ray observatories, including {\it Fermi } \citep{2022GCN.32636....1V,2022GCN.32642....1L},  {\it Swift} (for more detailed analysis see \cite{Williams:2023sfk}), SRG/ART-XC \citep{2022GCN.32663....1L}, Konus-Wind \citep{2022GCN.32668....1F} and others. The burst and its afterglow were also registered by  LHAASO on Earth in the range of hundreds  GeV to several TeV \citep{2022GCN.32677....1H}. Carpet-2 on Baksan Neutruno Observatory has detected an atmospheric shower 250\,TeV photon from the location of GRB 221009A  \citep{2022ATel15669....1D}. On the other hand HAWC collaboration reports no detection of photons from the afterglow in TeV range beyond  8 hours after the trigger \citep{2022GCN.32683....1A} and claim the upper limit on the energy flux $4.16\cdot 10^{-12}$\,TeV\,cm$^{-2}$\,s$^{-1}$.

 The burst was intrinsically strong and relatively nearby, z = 0.151.  The apparent brightness of GRB 221009A is exceptional. In {\it Fermi} GBM burst catalog it exceeds the next brightest by factor 15 in energy fluence. However the strongest impact of this event is due to claims of two photons 18\,GeV (LHAASO) and 250\,TeV (Carpet 2) which cannot come from z = 0.15 because of the absorption on extragalactic background light. There already have appeared numerous e-prints suggesting new physics to  explain these photons. 
 
Taking advantage of the brightness of GRB 221009A  we try to find something new about GRBs themselves in publicly available {\it Fermi} data.    Namely:
  
-- Is there anything interesting between the precursor of the burst and its main emission three minutes later.

-- What does the transition from the prompt phase of gamma-ray bursts to afterglow look like?

-- How bright is the afterglow and how long can it be traced in the GeV range.

\section{Data and their calibration}

\begin{figure}
	\includegraphics[width=\columnwidth]{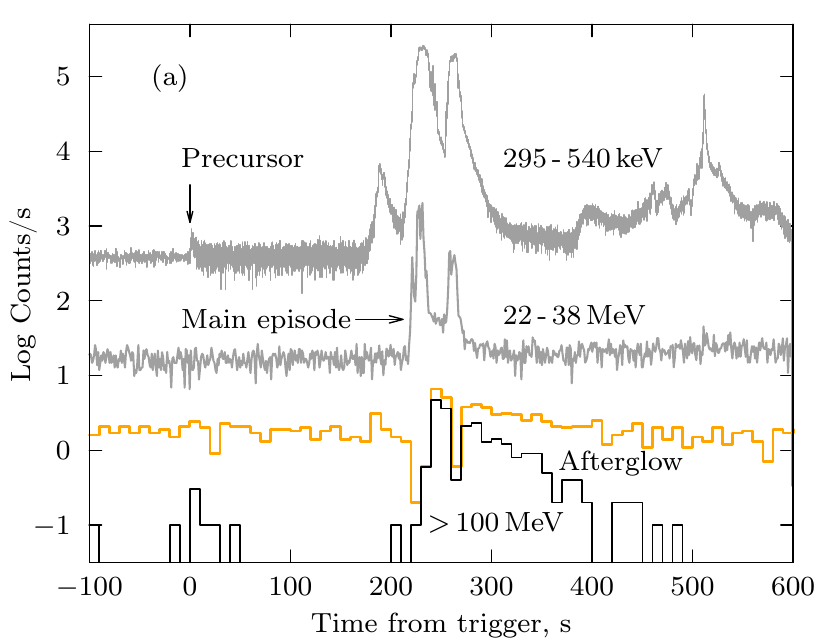}
	\includegraphics[width=\columnwidth]{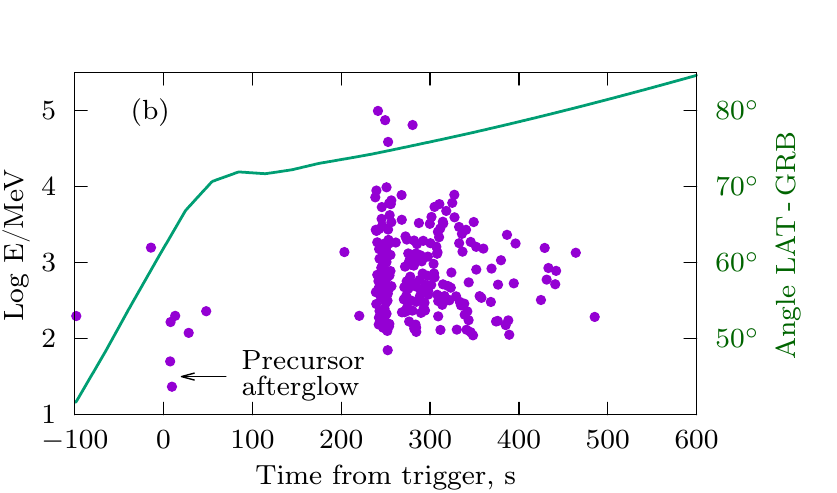}
    \caption{Time evolution of GRB 221009A in raw data. Prompt phase and early afterglow.  (a)\; Count rates in various energy ranges on a logarithmic scale, upper curve GBM NaI 295 - 540 keV, middle curve  GBM BGO 22 - 38 MeV,  lower blue curve  LAT, $8^\circ$ circle around the location of GRB 221009A, yellow curve - all LAT photons. Dips at 210 and 260\,s result from detector saturation. (b)\;  Individual LAT photons in the $8^\circ$ circle around the position of GRB 221009A (left logarithmic energy scale) and the angle between the LAT axis and the burst direction (right scale). Note the emission in the 10-200\,MeV range 50\,s after the precursor.}
\label{fig:fig1}
\end{figure}

Some raw {\it Fermi} data for GRB 221009A are shown in Fig.~\ref{fig:fig1}. Large Area Telescope (LAT) and NaI detectors of Gamma Burst Monitor  (GBM) were oversaturated while Bismuth Germanate (BGO) scintillation detectors satisfactory reproduce the peak flux in energy channels above $\sim$\,1\,MeV. There are no LAT data in the most interesting intervals 220 -- 240 and 260 -- 270 seconds (photon detections lack not only from the GRB direction but from the whole sky).

The next circumstance that makes the direct interpretation of LAT data problematic is a large angle $\theta$ between the direction of LAT z-axis and the burst location. The burst has occurred at the very edge of the telescope field of view where the detection efficiency is low.  Fig 1b shows time dependence of  $\theta$ on time during the event.  The inclination angle varied from $75^\circ$ to more than $80^\circ$ at $t \sim 500 $\,s when the source leaved the field of view for an hour.

The angular dependence of LAT effective area is given in  \cite{2021ApJS..256...12A}, see also 
\href{https://www.slac.stanford.edu/exp/glast/groups/canda/lat_Performance.htm}{Fermi LAT Performance}. 
However these data have insufficient resolution for this specific problem, therefore we have performed a detailed calibration of LAT detection efficiency for large incident angles using the brightest GeV  source Vela X (both the pulsar and the nebula). The same object was used by {Fermi}  team for calibration of the point spread function \citep{2021ApJS..256...12A} .
   
\begin{figure}
	\includegraphics[width=\columnwidth]{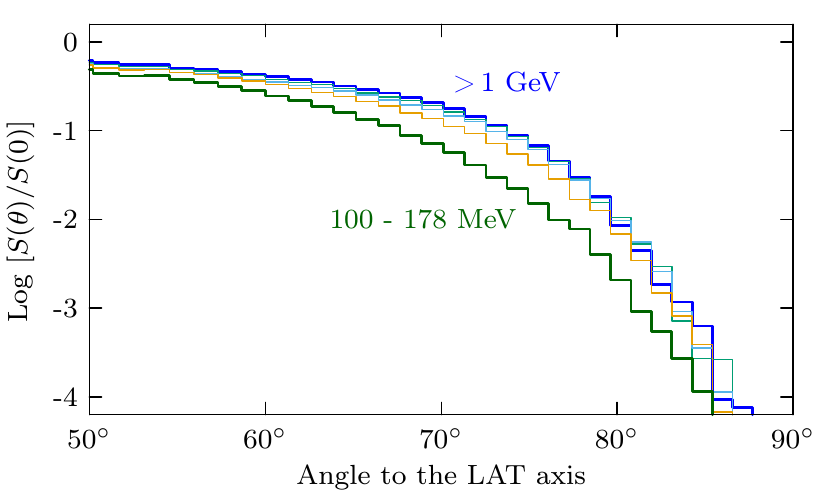}
    \caption{Calibration data for LAT performance at large incident angles using Vela-X gamma-pulsar.  The ratio of effective area to on-axis effective area as a function of angle between the photon source and LAT z-axis.  Curves from top to bottom correspond to energy intervals : $ > 1$ GeV, 1 GeV --  562 MeV, 562 MeV -- 316 MeV,  316 MeV -- 178 MeV, 178 MeV -- 100 MeV.}
\label{fig:fig2}
\end{figure}

We use photons in  $8^\circ$ circle around Vela X location as the calibration sample. The size of this circle is a result of a trade-off between sufficient containment for $\sim100$ MeV photons $( > 68\%)$ and the contamination of the sample with background photons. The result of our calibration is shown in Fig.~\ref{fig:fig2}. 

 The number of photons in our calibration sample is $9.4 \cdot 10^6$ while the number of background photons is $~ 2 \cdot 10^6$ as we have estimated from a neighbouring site of Milky Way. So, the background is considerable, but its effect should be moderate because the angular dispersion of background photons contributes the detection efficiency with different signs: the number of photons incident at $\theta + \Delta$ is comparable with those arrived at  $\theta - \Delta$. Neverphtheless wide angular cut in calibration sample mimics some extension of the field of view.  We have checked the effect of the wide angular cut with the same calibration cutting the sample at $4^\circ$. The difference at $\theta = 78^\circ$ for photons with $E  \gtrsim 1$ GeV is $~40\%$ and factor 2.5 for $100  - 178$ MeV interval. The latter large value is certainly the effect of widening of point spread function at edge of the field of view. Therefore   we prefer to use the calibration with $8^\circ$ calibration sample as it better reproduces the soft end of the spectrum and an extra background just slightly affects its hard range.

Except angular dependence of detection efficiency one should take into account the energy dependence. We use that described by \cite{2021ApJS..256...12A}.  Note that for lowest energy bin  100\,--\,178\,MeV that we use the efficiency is 0.45 of the maximal efficiency. Therefore a 100\,MeV photon detected by LAT at 400\,s after the trigger represents $\sim$\,300 photons of the same energy crossing the detector area (see Fig.~\ref{fig:fig1}b and Fig.~\ref{fig:fig2}). In our analysis and calibration we do not distinguish the conversion type and the quality class of photons, using total effective area for  the "source" event class.

\section{Main episode and onset of the afterglow}

\begin{figure} 
   \includegraphics[width=\columnwidth]{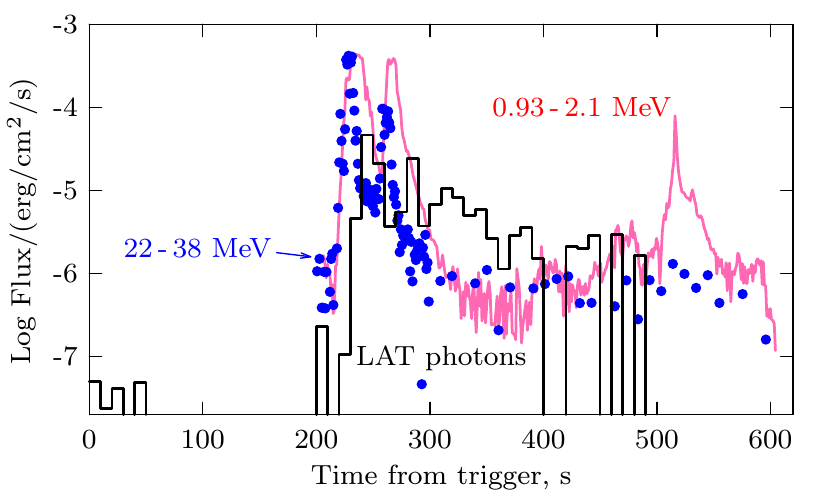}
    \caption{Prompt GRB and early afrerglow in different energy ranges. Histogram: energy flux of LAT photons normalized to angular-dependent effective area calibrated with Vela-X source.Green line: energy flux in 0.93\,-\,2.1\,MeV range from  Gamma-Burst Monitor BGO detectors. Blue circles: the same in 22\,-\,38\,MeV energy band.}
\label{fig:fig3}
\end{figure}

Normalized {\it Fermi} data for the first 600 seconds are shown in Fig.~\ref{fig:fig3}. LAT data were normalized using our Vela X calibration results (Fig.~\ref{fig:fig2}) and energy dependent effective area from \cite{2021ApJS..256...12A}. Total effective area of two  BGO detectors was set to 200 cm$^2$ independently on energy and incident angle as such assumption is sufficient for qualitative  demonstration. For the energy and angular performance of BGO detectors see \cite{Meegan_2009}.
 
BGO 22\,-\,38\,MeV energy flux in 300\,-\,600\,s interval is very sensitive to the background model. We use three parametric description: constant plus one sinusoidal half period with fitting intervals 300\,-\,150\,s and 600\,-\,1300\,s. Resulting $\chi^2$ is good, however the error in energy fluence in BGO us large therefore we do not take BGO data into consideration when reconstructing photon spectra.  We see a striking transition in time behaviour at $\sim$\,300\,s: a sharp decline of main pulse changes to a flat smooth slope. The exception is  another soft episode of prompt emission at 400 - 600 s which we discuss below. It would be reasonable to suggest then the high energy emission after 300\,s can be considered as the main afterglow. 

Phenomenologically,  prompt emission usually undergoes a very fast variation of  intensity and spectrum while afterglow has a smooth long decline with a stable wide spectral energy distribution. Prompt emission sometimes consists of multiple pulses of different duration and spectra, this pulses can overlap in time producing in some cases complex structures with a wide temporal Fourier power spectrum \citep{2000ApJ...535..158B}. Their time behaviour is very diverse. On the contrary, all afterglows have typical time behaviour: a power law decline slightly faster than $t^{-1}$. 
  
Theoretically, there exists a paradigm that the prompt emission arises from internal shocks (or magnetic reconnection or both) in the jet while the afterglow from external shock form collision of the jet with ambient medium (see e.g. \cite{RevModPhys.76.1143}), dominated by the stellar wind of the GRB progenitor \citep{Chevalier_1999}.  Physically, the prompt emission in many cases can be described as a radiation of an optically thick medium due to multiple Comptonization (e.g.~\cite{2018MNRAS.474.2828I}), while   afterglow better corresponds to synchrotron/Compton radiation of electrons accelerated in an optically thin environment, see \cite{2018IJMPD..2742003N} for a review.

\begin{figure}
   \includegraphics[width=\columnwidth]{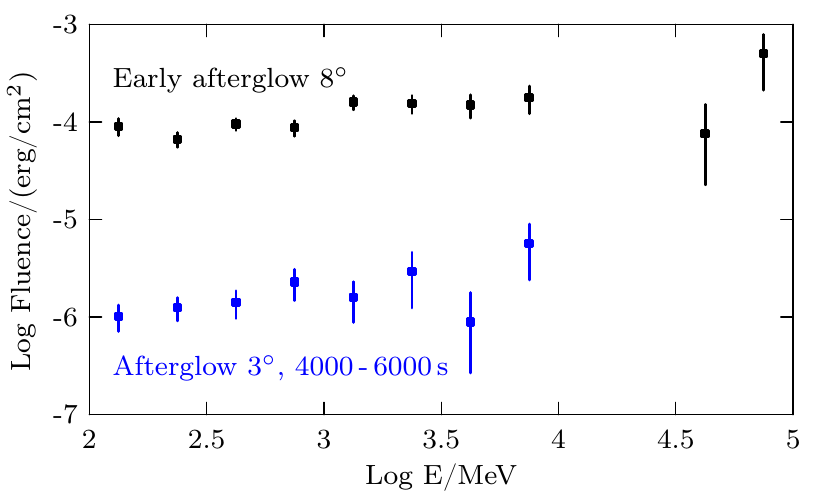}
   \caption{Spectral energy distribution of photons detected by LAT at time intervals 230\,-\,500\,s and 4000\,-\,6000\,s reconstructed with calibrated effective area (see Fig.~\ref{fig:fig2}). }
\label{fig:fig4}
\end{figure}

 In the case of  GRB 221009A we can describe as the prompt emission the precursor, a soft pulse at $t \sim 180\,-\,200$\,s, two main hard pulses and a soft long structure in $400\,-\,600$\,s interval. Photon flux detected by LAT since 250\,s with its spectrum (see Fig.~\ref{fig:fig4}) and light curve resembles an afterglow rather than a prompt  emission. The structure at 400\,-\,600\,s is probably an independent prompt emission episode overlapping in time with the early afterglow.

Probably the afterglow mechanism  (presumably the external shock) has turned on slightly earlier, e.g at 230\,s. Thereafter we use this time as a reference point for a power law decline of the afterglow. 
  
Our estimate of the energy fluence represented by 229 photons detected by LAT in the time interval 220 - 500\,s is $1.55\cdot 10^{-3}$\,erg/cm$^2$. Actual fluence could be several times higher since we don not know how many photons are lost in over-saturation gaps. The low energy fluence preliminary estimated by \cite{2022GCN.32642....1L} is $2.9 \cdot 10^{-2}$\,erg/cm$^2$, the total energy fluence can be much higher, see \cite{ 2023arXiv230213383F}. The spectral energy distribution in 100\,MeV - 100\,GeV range is shown in Fig.~\ref{fig:fig4}. We made a forward-folding power law fit to the numbers of detected photons in energy bins. The resulting photon index for LAT photons in background-free 8$^\circ$ field is -1.92$\pm 0.04$ which is consistent with  estimate of \cite{2022ATel15656....1P}  

\section{The main afterglow}

After 500 seconds since the trigger the GRB came out of LAT field. The next time window was open from $\sim 4100$ to $\sim 5700$ seconds. Then the location of GRB221009 periodically appeared in the field of view with duty cycle $\sim 20$ \%. 

\begin{figure}
   \includegraphics[width=\columnwidth]{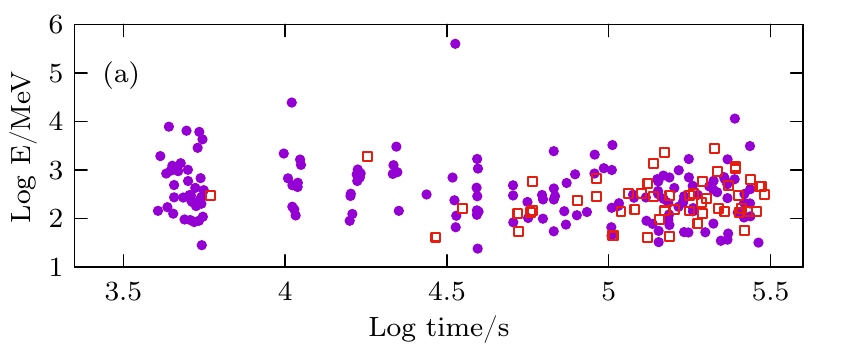}
   \includegraphics[width=\columnwidth]{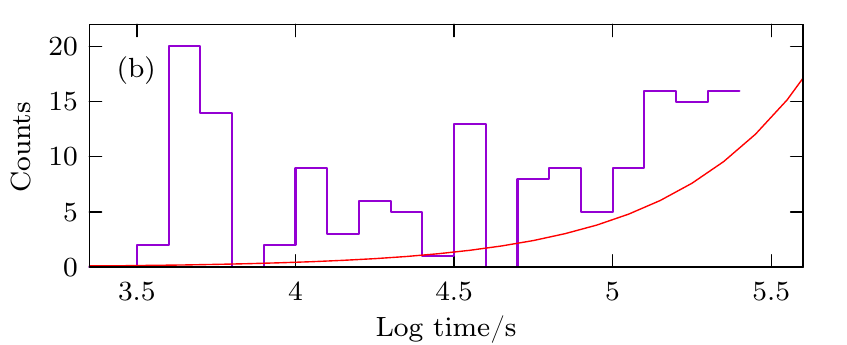}
\caption{Events in the  $1^\circ$ circle around GRB 221009A detected by LAT.  (a)\;Photons in the time interval $10^3$ - $3\cdot 10^5$ s since the trigger (green squares) and background photons in the interval of the same length but before trigger (red circles). 
 (b)\;The same but as a histogram. The average background normalized to the same logarithmic bins is shown by the green line. The estimate of the background has been done with $8^\circ$ field around the GRB location for 300000 seconds before the burst.}
\label{fig:fig5}
\end{figure}

Fig.~\ref{fig:fig5}  shows counts of photons detected by LAT versus logarithm of time. Unlike the main episode which is essentially background free, the background during the late afterglow is considerable and is too large in $8^\circ$ field of view which we accepted for the description of the main episode. For this reason we analyse the afterglow using $1^\circ$ circle. The background estimated with  photons detected from the same direction during 300 000 seconds prior the trigger is shown in Fig.~\ref{fig:fig5}. With this window the afterglow is significant up to two days (see Fig.~\ref{fig:fig5}b). Further contraction of the field of view does not improve the significance.
  
Fig.~\ref{fig:fig5}a shows LAT photons in log-log scale for $1^\circ$ field. It looks like the afterglow is getting softer with time, however note a 400 GeV photon at $\log(t) \sim 4.5$ (33554 s). The angular deviation of this photon from the GRB location is $0.06^\circ$, the probability of chance coincidence with such angular deviation during a day is $\sim 10^{-6}$. This photon is missing in the telegram of {\it Fermi} team, but was noticed by \cite{Xia-Zi-Qing}. The impression of softening can result from the lack of soft photons in the main episode and their excess at $t \sim 10^5$\,s. The former can be explained by a stronger suppression of soft photons at large incident angles (see Fig.~\ref{fig:fig2}a) and the latter by the contribution of a softer background. The statistics is insufficient to reveal an evolution of the afterglow spectrum. 
 
 \begin{figure}
   \includegraphics[width=\columnwidth]{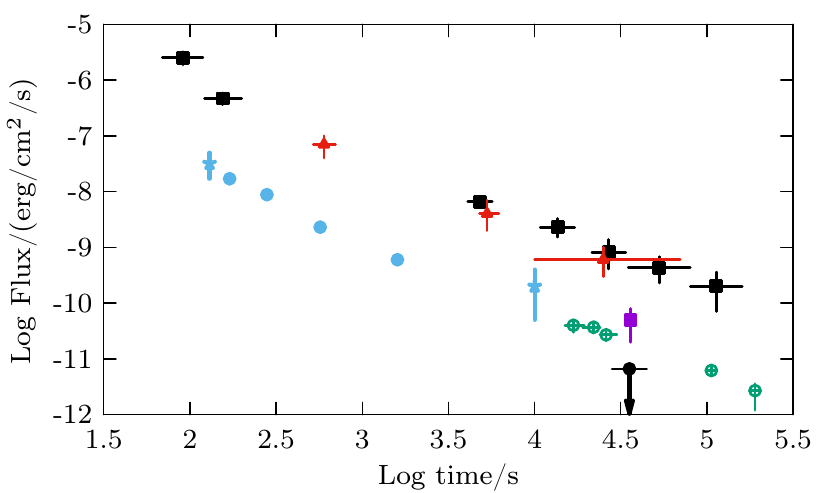}
    \caption{The afterglow of several $\gamma$-ray bursts. Black: GRB221009A, squares with error bars -- this work, dot with an upper limit -- HAWC Collaboration. Blue: GRB 190114C, dots -- MAGIC (0.3\,-\,1\,TeV), stars with error bars -- {\it Fermi} LAT. Green: GRB 190829A -- H.E.S.S. (0.2\,-\,4\,TeV). Violet: GRB 180720B -- H.E.S.S. (100\,-\,440\,GeV). Red: GRB 130427A -- {\it Fermi} LAT.}
\label{fig:fig6}
\end{figure}

The spectral energy distributions for the beginning of the afterglow and for the second window are shown in Fig.~\ref{fig:fig4}. They are consistent with each other and just slightly differ from the "canonical" spectrum with flat SED (photon index $\alpha = -2$). We suggest that it would be reasonable to set the spectral index to -2 when fitting the afterglow energy flux. Fig.~\ref{fig:fig6} shows the photon flux of the afterglow versus time. The flux is normalized to $3^\circ$ field of view using energy dependent relative containment in $3^\circ$ and $1^\circ$ circles averaged over spectrum with $\alpha = -2$.  We add for comparison some data points for high energy fluxes of other bright GRBs:  GRB 130427SA, \citep{doi:10.1126/science.1242353}, GRB 180720B~\citep{2019Natur.575..464A}, GRB 190114C~\citep{2019Natur.575..459M} and GRB 190829~\citep{HESS:2021dbz}. These data are summarised and discussed by \cite{ 2022Galax..10...66M}. The afterglow of GRB is slightly brighter than afterglow of GRB 130427A and almost an order of magnitude brighter than GRB 18072B , 190114C and 190829 which were detected by Cherenkov telescopes. This is the first case when the afterglow is still visible  in the Fermi range two days after the GRB.  Intriguingly, it was not detected by the HAWC Collaboration \citep{2022GCN.32683....1A},  see Fig.~\ref{fig:fig6}.  

The afterglow is still $3\sigma$ significant in 100 000 -- 200 000 s interval: the number of photons is 40 versus 23 photons in the "mirror" -100 000 -- -200000 s interval.  The afterglow light curve in 0.1 - 10\,GeV range can be described as a power law $F \sim t^{-\alpha_\tau}$, where $\alpha_\tau = 1.32\pm 0.05$.  

\section{The afterglow of the precursor}
\label{sec:maths} 

Fermi LAT detected 6 photons in the range 20 - 200 MeV in time interval 7 - 50 seconds after the trigger (Fig.~\ref{fig:fig1}b) when the angle $\theta$ varied from $65^\circ$  to $70^\circ$ and LAT effective area was several times less than for on-axis photons (see Fig.~\ref{fig:fig2}). Note that the prompt precursor emission already has relaxed when LAT has detected first of these photons.  We estimate the energy fluence represented by these 6 photons as $~1.5 \cdot 10^{-6}$\,erg\,cm$^{-2}$.

How significant is this "preafterglow"?  We have counted photons from $8^\circ$ circle when the orientation of LAT z-axis to the GRB was $63^\circ < \theta < 71^\circ$ during 30 000 seconds before the burst. The result is 302 photons for 23400 seconds which gives the expectation 0.64 photons for 50 s interval after the precursor. The probability to sample 6 photons by chance is $0.5\cdot 10^{-4}$ ($ 4 \sigma$).  Note that there is no "look elsewhere" effect in this case: the photons appear in a proper place with no sample manipulation.  Therefore this significance is quite sufficient to claim that {\it Fermi} has detected an afterglow of a precursor. This is the first case of such detection. The idea that a GRB precursor could produce its own afterglow was suggested by \cite{2014MNRAS.445.1625N}.

The spectral fit to 6 photons is, of course, quite loose, the resulting spectral index is $\alpha = -2.5\pm0.5$  which is consistent with the spectrum of the main afterglow with $\alpha \sim -2$.

\section{Discussion and conclusions}

Probably the most interesting fact that we see in {\it Fermi} data is the afterglow of the precursor. This is the first direct evidence of such phenomenon at least if we treat a precursor as a relatively weak event separated from the main episode by a long time interval. 

A weak precursor that occurs long before (up to several minutes) the main GRB is a feature of many GRBs. The estimates of their occurrence varies from 3\% GRBs \citep{1995ApJ...452..145K} up to 10\% \citep{2010ApJ...723.1711T} or even 20\% \citep{2005MNRAS.357..722L}. A useful review of the phenomenon is given by \cite{Zhu_PhD}.   In fact, this fraction may be even higher, since the precursor can be easily lost. The precursor of GRB 221009A has a few hundred times lower peak count rate and several thousand times lower energy fluence than the main emission episode. Such relatively weak precursor can by detected only in rare cases of very strong GRBs and one can not exclude that this phenomenon is common for the majority of GRBs  and we observe just strongest precursors. 

However this weak precursor has a bright afterglow. The energy fluence of the precursor and its afterglow is: $\sim 1.7\cdot 10^{-5}$\,erg\,cm$^{-2}$ (0.1 - 4.8\,MeV, BGO) and $\sim 1.5\cdot 10^{-5}$ erg\,cm$^{-2}$ correspondingly.  The estimate by \cite {2022GCN.32819....1M} of the precursor fluence is $2.38 \pm 0.04\cdot 10^{-5}$\,erg\,cm$^{-2}$. Therefore, they are comparable, while the main afterglow is much weaker than the main prompt emission episode:  $4.7\cdot  10^{-4}$\,erg\,cm$^{-2}$ versus $3.3\cdot  10^{-2}$\,erg\,cm$^{-2}$ (our estimate using BGO counts in 0.1 - 4.8\,MeV range), or even 0.21 erg cm$^{-2}$  in 0.02 - 10 MeV band according to estimate of \cite{ 2023arXiv230213383F}. The difference in this ratio is  two orders of magnitude. This is a hint that the precursor could differ from the main GRB in its nature. If we follow the paradigm that the prompt emission originates from internal shocks in the jet and the afterglow comes from external shock in the ambient medium (see \cite{RevModPhys.76.1143}), then we have to conclude that internal shock is pathologically weak in the case of the precursor. In principle, one cannot exclude that a precursor with its afterglow is an essentially  different phenomenon preceding the main emission. Unfortunately, there is a very small chance to observe such afterglow directly in the nearest future, nevertheless it probably could be sensed statistically using existing data bases.

As for the main afterglow, this is the brightest one as well as the GRB 221009 itself.  In other respects (relative energy flux, spectrum, decay law) it looks typical.  The afterglow due to its brightness is visible in {\it Fermi} data during two days and could be visible even longer by Cherenkov telescopes unless bad observational conditions including bright moon.  

HAWC collaboration has reported no detections beyond 8 hours after the event and set the upper limit $4.16\cdot10^{-12}$\,erg\,cm$^{-2}$\,s$^{-1}$ \citep{2022GCN.32683....1A}. This upper limit is in apparent tension with {\it Fermi} data unless one implies a sharp spectral decline between $\sim 10$\,GeV and TeV energy ranges. However other GRB do not have such decline in their afterglows which is supported by data of Cherenkov telescopes for other GRBs (cf. {\it Fermi} and MAGIC data in Fig.~\ref{fig:fig6}). Moreover, the 400\,GeV photon detected at 9 hours is an argument against a spectral cutoff. The publication of LHAASO results may clarify this issue.  Note that the second brightest afterglow of GRB 130427A also has not been detected above 100 GeV, the upper limit set by VERITAS \citep{ Aliu_2014} $\sim 10^{-11}$\,erg\,cm$^{-2}$\,s$^{-1}$  at the next day could be in some tension with {\it Fermi}  data too.

\section*{Acknowledgements}

We thank Fermi team for the excellent data base and NASA for the open data policy. 


\section*{Data Availability}

This work is based on publicly available {\it Fermi}  \href{https://fermi.gsfc.nasa.gov/ssc/data/}{database} and the quantitative   \href{https://www.slac.stanford.edu/exp/glast/groups/canda/lat_Performance.htm}{description}  of the {\it Fermi} Large Area Telescope performance. The work of  I.T. was supported by the Russian Science Foundation grant 23-42-00066.



\bibliographystyle{mnras}
\bibliography{grb}


\bsp	
\label{lastpage}
\end{document}